\documentclass[12pt]{article}
\usepackage{graphicx,float}
\usepackage{epsf,amsmath,bbold,amsfonts}
\usepackage{appendix}
\usepackage{amssymb}

\usepackage{xcolor}
\usepackage{hyperref}
\hypersetup{colorlinks=false,pdfborderstyle={/S/U/W 1}}

\def\b{\beta}
\def\c{\chi}

\def\D{\Delta}

\def\eps{\varepsilon}
\def\f{\frac}

\def\G{\Gamma}

\def\l{\left}

\def\la{\langle}
\def\ra{\rangle}

\def\mc{\mathcal}

\def\m{\mu}
\def\n{\nu}

\def\nn{\nonumber}

\def\p{\partial}

\def\r{\right}

\def\z{\zeta}

\def\be{\begin{equation}}
\def\ee{\end{equation}}

\def\bea{\begin{eqnarray}}
\def\eea{\end{eqnarray}}

\def\ba{\begin{array}}
\def\ea{\end{array}}

\def\bc{\begin{center}}
\def\ec{\end{center}}

\def\bl{\begin{flushleft}}
\def\el{\end{flushleft}}

\def\br{\begin{flushright}}
\def\er{\end{flushright}}

\def\bi{\begin{itemize}}
\def\ei{\end{itemize}}

\def\bt{\begin{tabular}}
\def\et{\end{tabular}}

\begin{document}




\begin{center}
{\bf \Large{ Comment on the paper "Minimal Fields of Canonical
Dimensionality are Free" by S.~Weinberg}}
\end{center}

\begin{center}
{\textsc {Alexander Monin, Mikhail Shaposhnikov}}
\end{center}

\begin{center}
 {\it 
Institut de Th\'eorie des  Ph\'eno\`menes Physiques, \\
\'Ecole
Polytechnique F\'ed\'erale de Lausanne, \\
Lausanne, CH-1015, Switzerland} 
\end{center}

\begin{center}
\texttt{\small alexander.monin@epfl.ch} \\
\texttt{\small mikhail.shaposhnikov@epfl.ch} \\
\end{center}

%


Recent publication \cite{Weinberg:2012cd} gives an elegant proof that in
a scale invariant relativistic field theory a field of canonical
dimensionality belonging to the minimal representation of the Lorentz
group is free.  Although the proof does not imply the existence of a Lagrangian - a theory can be given by a set of correlators (bootstrap) -  we will focus on the systems admitting Lagrangian formulation. In this case it is crucial for the proof to assume Lorentz and scale invariance of both the Lagrangian and the vacuum.  In this note we consider a somewhat weaker assumption. 
Taking a theory with scale invariance broken spontaneously and the Lorentz
invariance kept intact we demonstrate that minimal fields
of canonical dimensionality are not necessarily free. Briefly recapping the proof we show why the
arguments of \cite{Weinberg:2012cd} are not applicable for spontaneously
broken scale symmetry. 

Weinberg proposes to consider the action of the
operator
\be
{L} ^ \m_ { \n} = - i z ^\m \f {\p} {\p z ^ \n} + i z ^\n \f {\p} {\p
z ^ \m}
\ee
on the two-point function 
\be
G (x-y) = \la 0 \l |  \psi (x) \psi ^ \dagger (y)  \r | 0 \ra,
\ee
where $\psi$ is a field belonging to an arbitrary representation of
the Lorentz group. For brevity we suppress the index numbering the
components. Using the Lorentz invariance one gets
\be
L ^ \m _ \n L ^ \n _ \m G(z) = J^ {\m \n} J_ {\m \n} G(z)  - 2 J^ {\m
\n} G(z) J^ \dagger _ {\m \n} + G(z) J^ {\dagger \m \n} J^ \dagger _
{\m \n}.
\label{L_square}
\ee
where $J _ {\m \n}$ is the generator of the Lorentz transformations
for the representation of $\psi$. On the other hand a straightforward
computation yields
\be
L ^ \m _ \n L ^ \n _ \m = \l [ 2 S^2 -4 S - 2 z ^ 2 \Box \r ] G(z),
\label{L2}
\ee
with 
\be
S = - z ^ \m \f {\p} {\p z ^ \m}
\ee
being the generator of dilations.
The crucial assumption of the proof -- the scale invariance of the
vacuum -- translates into
\be
SG (z) = 2 \D G (z),
\label{no_spont_break}
\ee
with $\D$ being the scaling dimension of the field $\psi$. For the minimal field  (in the representation $(j,0)$ or $(0,j)$) the 
canonical dimensionality is $\D = j + 1$. As a result from (\ref{L_square}) and (\ref{L2}) we obtain the equation for the two-point function
\be
\Box \, G(z) = 0.
\ee
Therefore, one concludes that the field $\psi (x)$ is indeed free
\be
\Box _ x \psi (x) = 0.
\ee
Let us now turn to the case of spontaneously broken scale symmetry. If
the Lagrangian is invariant under the scale transformation the Ward
identities corresponding to the field transformation
\be
\delta \psi (x) = - \l (x ^ \m \p _ \m + \D \r ) \psi (x),
\label{scale_inf}
\ee
are not changed regardless whether the vacuum is invariant under the
symmetry or not
\be
i \la \p _ \m j _ D ^ \m (x) \psi (x _ 1) \dots \ra = \sum _ i \delta
(x - x _ i) \la \psi (x _ 1) \dots \delta \psi (x _ i) \dots \ra.
\label{Ward_id}
\ee
If the vacuum is invariant under the scale transformations, integrating
(\ref{Ward_id}) over the space-time for the two-point
function one gets precisely the formula (\ref{no_spont_break}). However, spontaneous
symmetry breaking renders the integral of the {\it l.h.s} of
(\ref{Ward_id}) non-zero. Usually in the case of spontaneous
symmetry breaking the vacua can be labeled by the vev $v$ of some
operator
\be
\la v \l | \mc{O} \r | v \ra= v,
\ee
which serves as an order parameter. In this case scale transformations
generated by (\ref{scale_inf}) relate the correlators computed over
different vacua. Namely, if the transformation is realized by the
unitary operator $U$
\bea
U \psi (x') U ^ \dagger & = & \lambda ^ {- \D} \psi (x), \nn \\
&& x'  =  \lambda x,
\eea
the expectation value becomes
\be
 v' = \lambda ^ {- \D _ {\mc{O}}} v.
\ee
While for the correlators one gets
\be
\la v ' \l | \psi (\lambda x _1) \dots \psi (\lambda x _N)  \r | v ' \ra = 
\lambda ^ {- N \D}
\la v \l | \psi (x _1) \dots \psi (x _N)  \r | v \ra.
\label{WI_sb}
\ee
The corollary is that for the case of spontaneous symmetry breaking
the formula (\ref{no_spont_break}) is not applicable and one cannot
conclude that the field is free. Below we give an example of 
an {\it effective} field theory with exact but spontaneously broken scale
symmetry which has massless interacting particle with canonical
scaling dimension.


We consider a toy model described by the following Lagrangian
\be
\mc{L} = \f {1} {2} \l ( \p _ \m \phi \r ) ^ 2 + \f {1} {2} \l ( \p _ \m \chi \r ) ^ 2 - \lambda _ 0 \l ( \phi ^ 2 - \z ^ 2 \chi ^ 2 \r ) ^ 2.
\label{Lagr}
\ee 
Classically it is scale invariant in $4$ dimensional space-time. The potential in (\ref{Lagr}) has a flat direction. Choosing the vacuum with non-zero vevs for the fields $\phi$ and $\chi$ breaks the symmetry spontaneously. As a result there are two interacting scalar particles in the spectrum. One of them is the Goldstone boson corresponding to the broken scale symmetry (dilaton), therefore, it is massless. The mass of the second particle is proportional to $\z \la \c \ra$.

However, this is not the end of the story. Although the symmetry is manifest at the classical level quantum corrections usually destroy the symmetry (see
for example \cite{Coleman_aspects,Brown:1979pq}), making the trace of
the energy-momentum tensor non-vanishing (we do not consider here
rather special case of theories with zero $\b$-function and, therefore, not
running coupling constant). Such an "anomaly" is usually attributed to the
regularization/renormalization procedure. Meaning that it is necessary
to introduce a mass parameter in one way or another which breaks the
symmetry explicitly (e.g., Pauli-Villars regulators have mass,
therefore, the symmetry is broken). That suggests the way out.
In \cite{Shaposhnikov:2008xi} it was proposed to use a modified version
of the dimensional regularization. The approach is somewhat analogous
to the one in \cite{Englert:1976ep}. It is not unique, another scale
invariant regularization was discussed in \cite{Shaposhnikov:2008ar}. 

Let us outline the idea. In the framework of standard dimensional regularization~(\cite{'tHooft:1972fi,Collins_book}) 
one considers the system in $n = 4 -\eps$ dimensions introducing the renormalized dimensionless coupling constant
\be
\lambda _ 0 = \lambda \m ^ {4-n} \l [ 1 + \sum _ {k} \f {C _ k} {(n-4) ^ k} \r ],
\ee
with $\m$ being an arbitrary renormalization scale needed to compensate the dimension of $\lambda_0$. The presence of this scale is the source of non-invariance. Therefore, promoting $\m$ to be field dependent 
\bea
\m & \to & \c ^ {\f {2} {n-2}} F_n (\phi/\chi),\nn \\
&&F_4 (\phi/\chi) = 1,
\eea
makes the scale symmetry manifest.

It was shown in \cite{Shaposhnikov:2008xi} that at one loop the  prescription described above indeed leads to the scale invariant effective potential. One can choose the counter terms needed to cancel the divergencies in such a way that the flat direction is preserved as well. That means that the scale symmetry is exact at the quantum level and spontaneously broken. As a result, the dilaton stays massless and interacting\footnote{It follows from the formula~(\ref{WI_sb}) or from the modified Lagrangian itself that the anomalous dimensions of the fields are zero.}. For energies much less than $\la \c \ra$ in the limit $\z \ll 1$ the momentum dependence of the matrix element $ \phi \phi \to \phi \phi$ (or equivalently of the $4$-point function $\G _ {4 \phi}$) coincides with the one prescribed by the standard renormalization group. Of course, the scale symmetry preserved at the quantum level with such a prescription is not given for free, new counter terms are needed at higher orders rendering the theory non-renormalizable~\cite{Shaposhnikov:2009nk}.

To conclude, in this note we considered the example of a theory with spontaneously broken scale invariance. The Goldstone boson of such a theory (dilaton), although having canonical dimensionality, {\it does not} have to be decoupled from other fields and the interactions can have long reaching phenomenological implications.

One of the reasons to study theories with spontaneously broken scale invariance originates from the desire to explain two puzzles, namely, the Higgs hierarchy and the cosmological constant problems. The scale symmetry in combination with approximate shift symmetry $\c \to \c + c$, existing at $\z \ll 1$ leads to stability of the Higgs mass against radiative corrections and to an alternative formulation of cosmological constant problem (for details see \cite{Shaposhnikov:2008xi,Shaposhnikov:2008xb}).

\section*{Acknowledgements}

We thank Sergey Sibiryakov for helpful discussions and valuable comments. This work was supported by the Swiss National Science Foundation.

\bibliographystyle{utphys}
\bibliography{free_fields_papers,free_fields_books}{}

\end{document}